\documentstyle[12pt,epsf]{article}
\renewcommand{\thefootnote}{\fnsymbol{footnote}}

\def\sst{\scriptscriptstyle}

\begin{document}

\leftline{SELF-AVOIDING RANDOM MANIFOLDS
\footnotemark[1]\footnotetext[1]{Lectures given at the Carg\`ese Summer School
``Low Dimensional Applications of Quantum Field Theory", July 1995.}
}
\vskip.65truein

\hbox{\obeylines\baselineskip12pt\parskip0pt\parindent0pt\hskip1.1truein
\vbox{Fran\c cois DAVID\footnotemark[2]
\vskip.1truein
Service de Physique Th\'eorique\footnotemark[3], CEA Saclay,
\vskip.1truein
F-91191 Gif-sur-Yvette CEDEX, FRANCE
}}
\vskip .65truein

\footnotetext[2]{Physique Th\'eorique CNRS}
\footnotetext[3]{Laboratoire de la Direction des Sciences de la Mati\`ere du CEA}
\renewcommand{\thefootnote}{\arabic{footnote}}

\section{Tethered Surfaces and Random Manifolds}
Several important developments of theoretical physics in the last 15 years
come from the extension of the concept of random walk to fluctuating
extended objects.
This has been very fruitful both in high energy physics, where the quantum
fluctuations of strings ($1+1$-dimensional objects) and of $p$-branes
($p+1$-dimensional)
in Minkowski space are considered, and in condensed matter physics,
where the thermal fluctuations of 2-dimensional films or membranes in
Euclidean 3-dimensional space are a fascinating subject
(see for instance \cite{r:Jerus}).

It is known that a 2-dimensional surface, with an intrinsic metric
$g_{\alpha\beta}$, embedded in flat $d$-dimensional
target space (the embedding being described by the mapping $x\to\vec r(x)$)
is characterized by its extrinsic metric
\begin{equation}
\label{eExMet}
h_{\alpha\beta}\ =\ \partial_\alpha\vec r\partial_\beta\vec r
\end{equation}
Dimensional analysis shows that the relevant terms involving $\vec r$ in the
action may involve one derivative $\partial_\alpha\vec r$ (tangent vectors)
and two derivatives $\partial_{\alpha\beta}\vec r$ (extrinsic
curvatures) of the embedded surface.
Two very different classes of models exist, characterized by the coupling
between the intrinsic and the extrinsic metric.
\par\noindent (1)
The intrinsic metric is proportional to the extrinsic one, i.e. is
coupled to the embedding by the constraint
$g_{\alpha\beta}=\partial_\alpha\vec r\partial_\beta\vec r$.
This is the case for the ``rigid string''
\index{string!rigid} model of Polyakov, and for the
equivalent Canham-Helfrich model of fluid membranes \index{membrane!fluid}
with bending rigidity.
The action is
\begin{equation}
\label{eCaHel}
S[\vec r]\ =\ \int d^2x \sqrt{g}\,\Big[\tau\,+\,{\kappa\over 2}\,
(\Delta\vec r)^2\Big]
\end{equation}
with $\tau$ the bare string (or surface) tension, and $\kappa$ the bending
rigidity.
Renormalization group calculations show that at large distances, the bending
rigidity becomes irrelevant, and that the effective action is nothing
but that of the Polyakov string, \index{string!Polyakov}
with an effective string tension $\tau_{\rm eff}$
dynamically generated
\cite{rPeLe85,rPol86}.
\par\noindent (2)
The intrinsic metric $g_{\alpha\beta}$ does not fluctuate (or it has
a dynamics decoupled from that of the extrinsic metric).
This class of models is not very useful for high energy physics, but is 
relevant in statistical physics to describe the so-called
``tethered membranes'', \index{membrane!tethered}
which generalize the concept of flexible 1-dimensional
chains (polymers) to 2-dimensional networks
\cite{rKaKN86}.
It is this class of models that I shall discuss in these lectures.

Let me consider a flexible two dimensional regular triangular network
fluctuating in 3-dimensional space.
A simple discrete action for the model is
\begin{equation}
\label{eDisAc}
S\ =\ -\,\kappa\!\sum_{{\rm neighbouring} \atop {\rm triangles}\ t, t'}\!
\vec n_t\cdot\vec n_{t'}\ +\ \tau\sum_{{\rm links}\ i,j} (\vec r_i-\vec r_j)^2
\end{equation}
where $\vec n_t$ is the normal vector to the triangle $t$.
The first term is the extrinsic curvature \index{curvature!extrinsic}
term, with $\kappa$ the
bending rigidity, the second term is a Gaussian elastic term.
Numerical simulations and analytical arguments indicate a very interesting
behavior for such a model.
If $\kappa$ is small enough, the bending rigidity is irrelevant at large
distances, and the surface is in a ``crumpled'', or ``collapsed'', phase,
with $\langle\vec n_t\rangle=0$.
This means that O(3) rotational invariance is not broken and that the surface
has no average orientation.
If $\kappa$ is large enough, the surface is in a flat phase, characterized by
a non-zero average orientation $\langle\vec n\rangle\ne 0$, and a spontaneous
breakdown of O(3) invariance.
These two phases are separated by a crumpling transition
\index{transition!crumpling} at some $\kappa_c$.
Numerical simulations indicate that (at $d=3$) this transition is continuous,
and characterized by non-trivial critical exponents.

An continuous model \`a la Landau-Ginzburg to describe
tethered surfaces is given by the effective action \cite{rPaKN88}
\begin{equation}
\label{eRigHam}
S\ =\ \int d^Dx\,\left[{\kappa\over 2}(\Delta\vec r)^2\,+\,{t\over 2}
\,h_{\alpha\alpha}\,+\,{K\over 2}(h_{\alpha\alpha})^2\,+\,\mu\,
\left(h_{\alpha\beta}-{\delta_{\alpha\beta}\over D}h_{\gamma\gamma}\right)^2
\right]
\end{equation}
with $h_{\alpha\beta}=\partial_\alpha\vec r\partial_\beta\vec r$ the
extrinsic metric.
$D$ is the internal dimension of the surface (the model describes in
fact $D$-dimensional tethered manifolds),
$\Delta=\partial_\alpha\partial_\alpha$ is the Laplacian,
$\kappa$ the bending rigidity. $t$, $K$ and $\mu$ are the effective elastic
moduli ($t$ corresponds to a tension, $K$ and $\mu$ are related to the so-called
Lam\'e coefficients).
Neglecting the fluctuations, the minimization of this action shows that
if $t>0$, $\langle\partial_\alpha\vec r\rangle=0$ and 
$\langle h_{\alpha\beta}\rangle =0$, so that the surface is crumpled, while
if $t<0$ $\langle \partial_\alpha\vec r\rangle\ne 0$ and the surface is flat.
$\langle h_{\alpha\beta}\rangle$ vanishes and there is a continuous
crumpling transition at $t=0$.
This analysis neglects the effect of fluctuations and is valid only if
the internal dimension of the manifold is $D>4$.
For $D<4$ fluctuations become important, and their effect
can be estimated by an $\epsilon$-expansion for $D=4-\epsilon$
\cite{rPaKN88}, or by a large $d$ expansion 
\cite{rDaGu88,rArLu88,rPaKa89}
(where $d$ is the dimension of the target space).
It might seems surprising that for 2-dimensional manifolds ($D=2$)
the flat phase
phase still exists, since it is characterized by a spontaneous breakdown
of the continuous rotational O($d$) symmetry, which should be forbidden by the
Mermin-Wagner-Coleman theorem.
Such a crumpling transition is indeed forbidden for $D=1$ (polymers):
for any dimension $d\ge 1$ of target space, infinite semi-flexible polymers
with a non-zero bending rigidity $\kappa$ are always crumpled at large
distance, and the flat phase does not exist.
In fact there is no contradiction for $D=2$.
The transverse degrees of freedom of the manifolds $\vec r_\perp$ (undulations)
are coupled to the longitudinal degrees of freedom $\vec r_\|$ (phonons),
so that the global symmetry of the model is not the compact group ${\rm O}(d)$
(rotations), but the non-compact group of Euclidean displacements ${\rm E}(d)$
(translations+rotations).
There is a non-trivial coupling between phonons (longitudinal modes) and
undulations (transverse modes) which generates effective long range interactions
between these transverse modes. 
In the presence of such long range interactions the Mermin-Wagner theorem
does not apply.

Let me give a tentative picture which emerge from the analytical and numerical
studies of the crumpling transition.
For $D>4$ the crumpling transition is continuous and its critical exponents
are given by mean field theory.
For $D=4-\epsilon$, $\epsilon$ small, one loop calculations indicates that
the crumpling transition is continuous for $d>d_c$ large enough, but becomes
discontinuous (fluctuation induced first-order) for $d<d_c$ small.
Large $d$ calculations shows that there is a second order crumpling transition
for $D\ge D_1(d)< 2$ with $D_1(\infty)=2$.
For $D<D_1(d)$ there is no flat phase and no crumpling transition.
Thus we expect that the domain $1\ge D\ge 4$ in the $(d,D)$ plane will be
separated into three regions.
A domain {\bf A} for small $D$ where the manifold is always crumpled,
a domain {\bf B} where there is a second order crumpling transition,
and a domain {\bf C} where the crumpling transition is always first order.
Numerical simulations indicates that the point $(D=2, d=3)$ is in {\bf B}
\cite{rKaNe87}.
Recent studies of the folding problem show that the point $(D=2,d=2)$ is in
{\bf C} \cite{rDFGu94}.
\begin{figure}
\centerline{\epsfxsize=10.cm\epsfbox{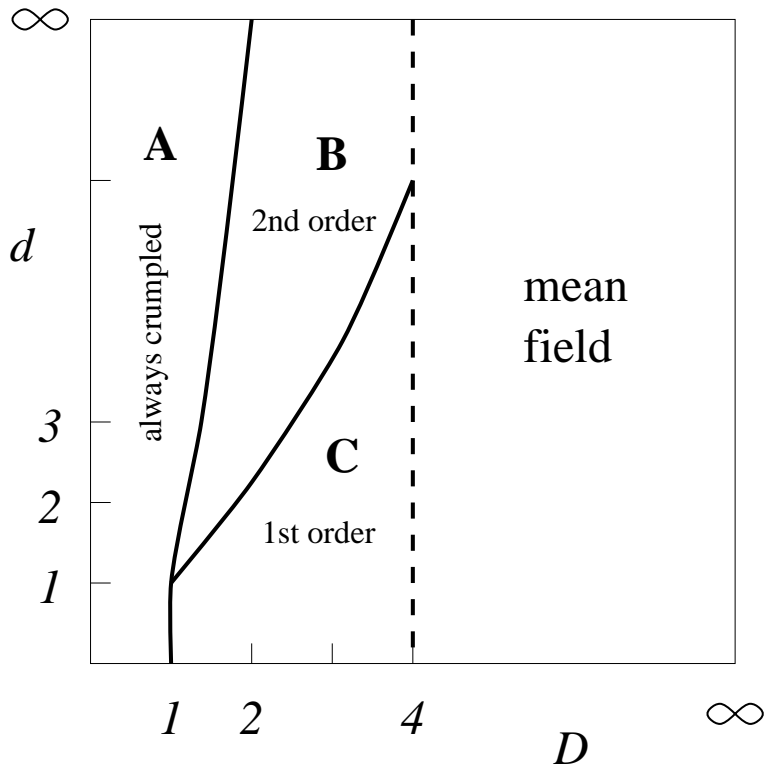}}
\caption{Nature of the crumpling transition for phantom manifolds
as a function of $D$ and $d$}
\label{fCrumTr}
\end{figure}

The crumpled phase is the simplest to characterize.
It corresponds to $t>0$ in (\ref{eRigHam}), and at large distance only the
quadratic $t$ term is relevant.
Therefore the action for a crumpled manifold 
\index{manifold!crumpled}
is Gaussian, and is nothing but the massless free field action
\begin{equation}
\label{eGauAct}
S\ =\ \int d^Dx\,{1\over 2}(\nabla_x\vec r)^2
\end{equation}
The properties of such manifolds are easy to compute.
For instance, if one considers a finite manifold with an internal extent $L$,
its average squared size $\langle R^2\rangle$ in target space scales as
\begin{equation}
\label{eScGau}
\langle R^2\rangle\  \propto\ \cases{\ L^{(2-D)}&\hbox{\ if $D<2$}\cr
\ \ln(L)&\hbox{\ if $D=2$}\cr
\ \rm{constant}&\hbox{\ if $D>2$}}
\end{equation}
which implies that the fractal dimension of a crumpled Gaussian manifold is
$d_f=2D/(2-D)$ if $D<2$, and is infinite if $D\ge 2$.

\section{Self-avoiding crumpled Manifolds}

The above considerations apply to ``phantom manifolds'',
\index{manifold!phantom}
 which are free to intersect themselves.
Indeed, the action (\ref{eRigHam}) takes into account only local couplings in the internal
space.
Such local couplings are the only relevant one for strings, but for
physical tethered networks self-avoiding interactions, 
\index{manifold!self-avoiding}
which involve elements of the manifold which are close in target space but are
arbitrarily far apart in internal space, are relevant.
It is expected that such interactions will change the scaling properties
of the manifold in the crumpled phase and at the crumpling transition.
For instance, in the crumpled phase, the average squared size will now scale as
\begin{equation}
\langle R^2\rangle\ \propto\ L^{2\nu}\qquad 0<\nu<1
\label{eNu}
\end{equation}
with a critical exponent $\nu\ge\nu_0={\rm sup}[(2-D)/2, 0]$.
If $\nu_0<\nu<1$ the manifold will be swollen by self-avoidance, but still
crumpled.
This is the case for polymers $D=1$ for $1<d<4$, with for instance $\nu=3/4$
for $d=2$.
The effect of self-avoidance may even be so strong that the manifold
stays flat and that the crumpling transition disappear, in this case
$\nu=1$.
Such a behavior has been observed in several numerical simulations of 
self-avoiding tethered membranes in three dimensions.

A simple analytical model to describe crumpled self-avoiding manifolds has been
introduced in \cite{r:AroLub87,r:KarNel87}.
It is a simple extension of the continuous Edwards model for polymers.
The action $S$ (the free energy for a configuration $\vec r$) is the sum of a
Gaussian
elastic energy and of a 2-body repulsive interaction, proportional to the
coupling constant $b$:
\begin{equation}
\label{eHam}
S[\vec r]\ =\ \int d^D\! x\ {1\over 2} (\nabla_{\!x}\vec r) ^2\ +\
b\ \int d^D\!x\int d^D\!y\ \delta^{d}(\vec r(x)-\vec r(y))
\ .
\end{equation}
The internal dimension $D$ may be taken as a continuous parameter,
interpolating between polymers ($D=1$) and membranes ($D=2$).
The issue is to compute the critical exponents describing the
scaling behavior of large manifolds, for instance the exponent $\nu$
(related to the fractal dimension $d_f$ of the manifold by $\nu=D/d_f$), and the
configuration exponent $\gamma$, related to the scaling of the partition
function $Z$ of a finite manifold with internal extent $L$ by
\begin{equation}
Z\ \propto\ L^{\gamma-1}\ \hbox{\rm constant}^{L^D}\ .
\label{eGamma}
\end{equation}
The mean field exponents are obtained by setting $b=0$.
One recovers the free Gaussian action and the exponents
$\nu_{0}=(2-D)/2$ and $\gamma_{0}=1-d(2-D)/2$ (if $D$ is not integer).

Dimensional analysis shows that the mean field theory is invalid if 
the engineering dimension of $b$, $\epsilon$, is positive
\begin{equation}
[b]\ =\ \epsilon\ =\ 2D-d(2-D)/2\ >\ 0\ .
\end{equation}
\begin{figure}
\centerline{\epsfbox{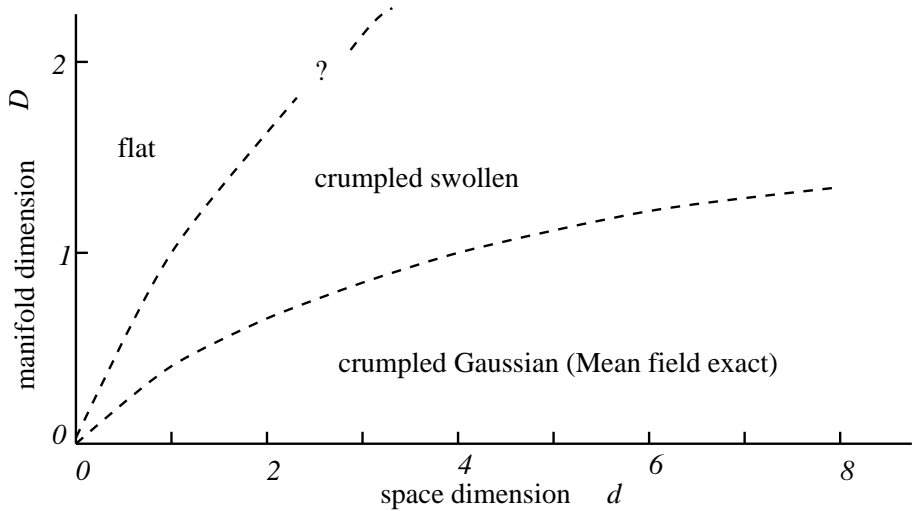}}
\caption{Self-avoiding manifolds as a function of $d$ and $D$}
\label{fdDplane}
\end{figure}
In this case, we expect that $\nu>\nu_0$.
For small enough $d$, and certainly for $d\le D$, we expect that the manifold
is flat.
The general picture of the expected behavior as a function of the internal
dimension $D$ and of the external dimension $d$ is presented in Fig.
\ref{fdDplane}

A natural idea is to compute the corrections to mean field by a
$\epsilon$-expansion \`a la Wilson-Fisher.
This has been done for polymers, first by using the so-called de Gennes trick
\cite{r:PGG72}:
in the scaling limit the self-avoiding walk can be mapped (by a Legendre
transform) onto a local field theory with O($n$) symmetry defined in the  
$d$-dimensional target space, in the limit $n\to 0$.
It is then equivalent to study the limit of a single very long polymer
and the massless limit of the $n=0$ theory, for which
standard renormalization group theory is appliquable.
Unfortunately no such equivalence exists for manifolds, beyond the case $D=1$. 
Another renormalization scheme used for polymers is the so-called direct
renormalization scheme \cite{r:Clo81}.
\index{renormalization!direct}
It has been applied to self-avoiding manifolds by Aronowitz \& Lubensky
\cite{r:AroLub87}\ and by Kardar \& Nelson \cite{r:KarNel87}, who 
performed calculations to first order in $\epsilon$.
The basic idea of this method is to perform explicit perturbative calculations
for a finite manifold and for $\epsilon>0$.
Perturbation theory is then UV and IR finite, but has UV divergences when
$\epsilon\to 0$.
It appears that these poles in $1/\epsilon$ can be removed by reexpressing the
observables in
terms of adequate dimensionless renormalized quantities, such as the
second virial coefficient.
The internal size $L$ of the manifold plays the role of the inverse of
a renormalization 
mass scale, and renormalization group equations can be obtained
by considering the $L$ dependence of the renormalized theory.
At first order in perturbation theory, the consistency of these renormalization
group equations has been checked explicitly by Duplantier, Hwa \& Kardar
\cite{r:Dup&al90}.
However, at that time it was not clear whether this direct renormalization
approach could be justified beyond first order (except for $D=1$, where
the de Gennes trick is used to show the equivalence between direct
renormalization and the standard minimal subtraction scheme).

\section{Renormalization for multi-local Theories:}
Recently is became possible to prove the consistency of this approach, and
the renormalizability of the model (\ref{eHam}) directly in the internal
$D$-dimensional space, despites the fact that this model is a non-local field
theory in $D$ dimensions \cite{r:DDG3}.
\index{renormalization!multi-local theories}
Let me present the general idea for the proof, which is due to B.~Duplantier,
E.~Guitter and myself.
The perturbation theory for this model is obtained by expanding the observables
as power series in $b$.
The bi-local ``interaction vertex" (in field theoretic language I call it a
bilocal operator) is written in Fourier transform as
\begin{equation}
\raisebox{-3.ex}{\epsfbox{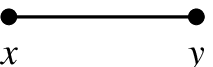}}\ \ =\ \delta^{d}(\vec r(x)-\vec r(y))\ =\ \int d^d\!k\ 
{\rm e}^{i\vec k(\vec r(x)-\vec r(y))}\ .
\label{eBilOp}
\end{equation}
It can be viewed in a Coulomb gas representation as the integral over the
``charge" $\vec k$ of a neutral ``dipole" with charge $+\vec k$ at $x$ and
charge $-\vec k$ at $y$.
The term of order $b^K$ in the perturbative expansion of the partition function
(as well as of other observables) involves $K$ dipoles
$(x_1,y_1),\ldots,(x_K,y_K)$.
The integration over the charges $\vec k_1,\ldots\,\vec k_K$ gives an integral
over the positions of the dipoles of the determinant of the ``dipole energy"
quadratic form $Q$
\begin{equation}
\int\cdots\int \prod_{i=1}^{K}\ d^D\!x_i\,d^D\!y_i\
\det\Big[Q[x_i,y_i]\Big]^{-d/2}\ .
\end{equation}
$Q$ is a $K\times K$ matrix such that $\sum\limits_{i,j=1}^{K}\vec k_i
Q_{ij}\vec k_j$ is the Coulomb energy (in $D$-dimensions) of the $K$ dipoles.
Each $Q_{ij}$ is a linear combination 
\begin{equation}
Q_{ij}\ =\ G_0(x_i,x_j)+G_0(y_i,y_j)-G_0(x_i,y_j)-G_0(x_j,y_i)\ .
\label{eGzero}
\end{equation}
of the Coulomb potentials $G_0$ between the endpoints of pairs of dipoles $i$ and $j$
\begin{equation}
G_0(x,x')\ =\ \left.\langle r(x)r(x')\rangle\right._{\!0}\ =
 {\Gamma\big((D-2)/2\big)\over 4\pi^{D/2}}\  |x-x'|^{2-D}
\ \ .
\label{eCoulPot} 
\end{equation}
This Coulomb potential is properly defined for $0<D<2$ by analytic continuation
in $D$.
For $0<D<2$ it is negative, but it vanishes for $x=x'$ (while for $D>2$ it
diverges), and it decreases at large distances (as for $D>2$).
The integration over the $2K$ points in a non-integer $D$-dimensional space can
also be defined properly by analytic continuation in $D$ and the use of
distance geometry.
This amounts to replace the integration over the $2K\times D$
coordinates of the $2K$ points by an integration over the $K\times (2K-1)$
scalar distances between these points.

One can show that when the determinant $\det[Q]$ vanishes short distance UV
singularities occur in the integrals.
This occurs if and only if some of the end-points of (not necessarily the same)
dipoles coincide, so that the end-points form ``atoms", while the dipoles form
``molecules'', and if moreover one can assign non-zero charges $\vec k_i$ to
the dipoles while each atom stays globally neutral. 
This condition is more easily depicted graphically on Figure 3.
\begin{figure}
\centerline{\epsfbox{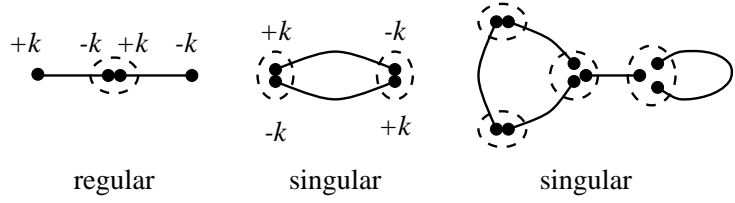}}
\caption{A non-neutral regular configuration and two neutral UV singular
configurations}
\end{figure}

The associated singularities of these integrals are related to the
behavior at short distance of the expectation value (with respects to the free
Gaussian model) of products of bilocal operators as given by Equ.~\ref{eBilOp}.
One can show that this short distance behavior is encoded in a multilocal
operator product expansion (MOPE)
\index{operator product expansion!multilocal}, which generalizes Wilson's
operator product expansion\index{operator product expansion}.
Let me give two examples:

When the two points $x$ and $y$ of the bi-local interaction operator tend
towards a single point, this operator can be expanded in terms of local
operators involving derivatives of the field $\vec r$.
The first terms of the expansion are explicitly (not writing explicitly
the $D$ and $d$ dependence of the coefficients)
\begin{eqnarray}
&&\vbox{\vskip 30pt}\nonumber\\
\ \ &=\ \hphantom{+\ }&c_0\ |x-y|^{\epsilon-2D}\ {\bf 1}\nonumber \\
\raisebox{-0.ex}{\epsfbox[0 0 27 0]{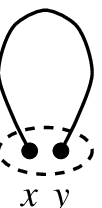}}
&& +\ c_1\ |x-y|^{\epsilon-D-2}(x^\alpha-y^\alpha)(x^\beta-y^\beta)\
:\nabla_{\!\alpha}\vec r\,\nabla_{\!\beta}\vec r:\nonumber \\
&&+\ \cdots
\label{eMOPE1}
\end{eqnarray}
${\bf 1}$ is the identity operator (its expectation value is 1),
the $:\ :$ in the operator $:\nabla_{\!\alpha}\vec r\,\nabla_{\!\beta}\vec r:$
denotes the normal ordering subtraction prescription required to deal properly
with the UV singularities contained in $\nabla_{\!\alpha}\vec
r\,\nabla_{\!\beta}\vec r$.

The second example is less simple, and shows that when the end-points of two
bilocal operators tend pairwise towards two different points, this generates
again bilocal operators
\begin{eqnarray}
&=\ &d_0\ \left[|x_1-x_2|^{2-D}+|y_1-y_2|^{2-D}\right]^{-d/2}\ 
\epsfbox{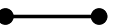}
\nonumber \\
\raisebox{-0.ex}{\epsfbox[0 0 92 0]{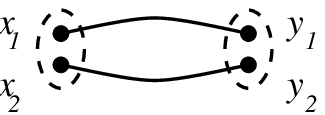}}\ 
&&\ +\ \cdots
\end{eqnarray}

This structure is generic, and products of local and bilocal operators generate
multilocal operators of the general form
\begin{equation}
\Phi\{x_1,\cdots,x_P\}\ =\ \int d^d\vec r_0\ \prod_{i=1}^P\
\left[\left(\nabla_{\vec r_0}\right)^{m_i}\ \delta^d(\vec r_0-\vec r(x_i))\
A_i(x_i)\right]\ .
\label{eMulOp} 
\end{equation}
where the $A_i(x_i)$ are local operators, which can be decomposed into
products of multiple $x$-derivatives of $\vec r$.
The $m_i$ are integers.
For $P=1$ and $m=0$ one recovers local operators $A(x)$ ($m>0$ gives $0$).
For $P=2$, $m_1=m_2=0$ and $A_1=A_2={\bf 1}$ one recovers the bilocal
interaction operator, etc$\ldots$ 
These operators have a very special form: they can be viewed as a local
convolution in the target $d$-dimensional  $\vec r$ space of a non-local
product (in the internal $D$-dimensional space) of the $P$ local operators
$A_i$.

The MOPE implies that the formalism of renormalization theory and of
renormalization group equations, which has been developed for local quantum
field theories, can be adapted for this model.
One is in fact interested in the IR scaling behavior of the lattice model,
when some length scale $L$ goes to $\infty$.
This lattice model is described by the action (\ref{eHam}), with a
short distance lattice cut-off $a$.
To study this IR limit it is equivalent to look at the UV continuum limit of
the model when the physical length scale $L$ is kept fixed, while the UV
cut-off $a$ goes to $0$.
In this limit one can construct, via renormalization, a finite renormalized
theory with $a=0$, which obeys renormalization group equations.
From these equations, one recovers the large distance behavior of the
lattice model we started from.
The procedure works well in perturbation theory when one is close to the upper
critical dimension, i.e. for $\epsilon$ small, and it
leads to the $\epsilon$-expansion.

\begin{figure}
\centerline{
 \hbox{
  \epsfbox{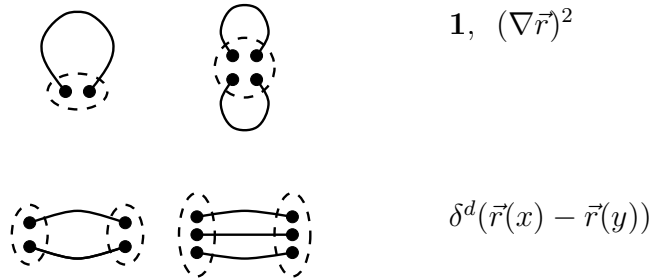}\kern10.ex
  \vbox{
   \hbox{{\bf 1},\ \ $(\nabla\vec r)^2$}
   \hbox{\vspace{8.5ex}}
   \hbox{$\delta^d\big(\vec r(x)-\vec r(y)\big)$}
   \hbox{\vspace{1.ex}}
  }
 }
}
\caption{UV divergent configurations and the associated relevant operators}
\label{fDiv}
\end{figure}
In our case, the MOPE can be used to determine, by power counting, which
multilocal operators are relevant and give UV singularities (poles in
$1/\epsilon$).
Then one can also show that these poles can be subtracted by adding to the
action (\ref{eHam}) counterterms proportional to the marginally
relevant multilocal operators, leading to the UV finite renormalized theory.
For the model of self-avoiding manifolds, this analysis shows that the
UV divergences
are associated only with local and bilocal operators,  as depicted on
Fig.~\ref{fDiv}, and that only three
operators are relevant:
the identity operator ${\bf 1}$, the elastic energy operator ($\nabla\vec r)^2$
and the bilocal operator $\delta^d(\vec r(x)-\vec r(y))$.
${\bf 1}$ is strongly relevant, and gives power-like UV divergences
proportional to $a^{-D}$ ($a$ being a short-distance cut-off).
The two other operators are superficially relevant, they give logarithmic
UV divergences or equivalently poles in $1/\epsilon$ at $\epsilon=0$.

\begin{figure}
\centerline{\epsfbox{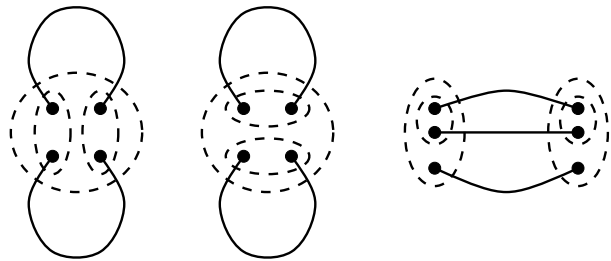}}
\label{fNesDiv}
\caption{Examples of nested singular configurations}
\end{figure}

The fact that the so-called superficial divergences, associated to a global
contraction of points towards a singular configuration, can be subtracted by
counterterms is a  consequence of the MOPE.
A complete proof of the renormalizability of the theory is possible, but much
more delicate.
It requires a control of the subdivergences coming from successive contractions
associated to nested singular configurations, such as those depicted of
Fig.~\ref{fNesDiv} .

\section{Scaling for infinite self-avoiding Manifold}
A first application of this formalism is the derivation of scaling laws.
Since the model is renormalizable (at least perturbatively),
it can be made UV finite (for $\epsilon\simeq 0$) by introducing two
counterterms in the action.
The new renormalized action is of the form
\begin{equation}
S[\vec r]\ =\ {Z\over 2}\,\int d^D\!x\ (\nabla \vec r)^2\ +\ b_R\,\mu^\epsilon\ 
Z_b\ \int\int d^D\!x\,d^D\!y\ \delta\big(\vec r(x)-\vec r(y)\big)\ .
\label{eRenHam}
\end{equation}
$b_R$ is the dimensionless renormalized coupling constant (the perturbative
expansion in $b_R$ is UV finite order by order).
$Z$ is a wave-function renormalization factor and $Z_b$ a coupling constant
renormalization factor, both are perturbative series in $b_R$, with poles up to
degree $1/\epsilon^{K-1}$ at order $K$.
$\mu$ is the renormalization momentum scale.
As for ordinary local theories, such as the Landau-Ginzburg-Wilson $\Phi^4$
model, one can change $b_R$ and $\vec r$ in Equ.~\ref{eRenHam} into bare
quantities in order to rewrite the renormalized Hamiltonian as a bare action
given by Equ.~\ref{eHam}.
The renormalization group $\beta$-function and the anomalous dimension $\gamma$
of the field $\vec r$ are defined in the standard way
\index{renormalization!group}
\begin{equation}
\beta(b_R)\ =\
\left.\mu{\partial\over\partial\mu}b_R\right |_{\sst\rm bare}
\qquad;\qquad
\gamma(b_R)\ =\ -\,{1\over 2}
\left.\mu{\partial\over\partial\mu}\ln
Z\right |_{\sst\rm bare}\ .
\end{equation}
\begin{figure}
\centerline{\epsfbox{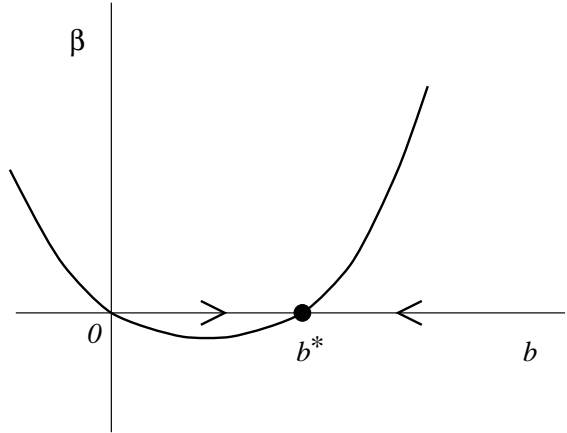}}
\label{fBeta}
\caption{The $\beta$-function and the RG flow for $\epsilon>0$.}
\end{figure}

The $\beta$-function is found to be of the form
\begin{equation}
\beta(b_R)\ =\ -\epsilon\, b_R+\hbox{\bf c}\,b_R+{\cal O}(b_R^2)
\qquad;\qquad\hbox{{\bf c}\ =\ {\bf c}(D) positive constant}\ ,
\label{eBeta}
\end{equation}
and therefore there is, at least for small $\epsilon>0$, an IR attractive 
fixed point $b_r^\star={\cal O}(\epsilon)$, which governs the scaling behavior
of self-avoiding polymerized surfaces at large distance.
The existence of this fixed point ensures the universality of this
non-trivial scaling for $\epsilon>0$, and that
no new interactions, possibly non-local in
external space, are generated by the RG transformations.

The explicit calculation for the scaling exponents $\nu$ and
$\gamma$ leads to the same results for the scaling exponents $\nu$ and
$\gamma$ than the direct renormalization method at first order in $\epsilon$.
With this method higher order calculations are feasible, but
technically quite difficult.
In particular, already at second order the RG functions cannot be expressed
analytically, and numerical integration methods have to be developed.
Work is in progress to compute the scaling exponents at order $\epsilon^2$.

\section{Finite Size Scaling and direct Renormalization}
The model given by Equ.~\ref{eHam} describes an infinite manifold with flat
internal metric, corresponding to an infinite and regular flexible lattice.
Finite manifolds are described by a similar model, but the $D$-dimensional
manifold $M$ is now embodied with a fixed non-trivial Riemannian metric 
$g_{\alpha\beta}(x)$ (examples are closed manifolds with the
topology of the sphere ${\cal S}_D$ or the torus ${\cal T}^D$), and may have a
boundary $\partial M$ (open manifold with the topology of the disk for
instance).
A similar analysis can be performed for such models, and the MOPE structure of
short distance singularities is still valid, but new local operators $A(x)$,
which depend on the internal metric on $M$ and on the boundary $\partial M$,
such as the scalar curvature 
\index{curvature!scalar}
$R$, appear in the MOPE and in Equ.~(\ref{eMulOp}).
The renormalized action now contains at least five operators and five
independent renormalization factors $Z$
\begin{eqnarray}
S[\vec r]\ =\ &&\int_M\,Z\,\hbox{\bf 1}\ +\ \int_M\,Z\,(\nabla\vec
r)^2\ +\ \int\!\int_M\,Z\,b\,\delta^d(\vec r-\vec r)\nonumber\\
&&+\ \int_M\,Z\,R\ +\ \int_{\partial M}Z\,\hbox{\bf 1}\ \ .
\label{fRenHam2}
\end{eqnarray}
The curvature operator $\int_M R$ is superficially relevant only for $D=2$ and
the boundary operator $\int_{\partial M}\hbox{\bf 1}$ only for $D=1$.
When these additional terms are not relevant, the first three renormalization
factors $Z$ are the same for finite curved manifolds than for the infinite flat
plane.
This property is analogous to the renormalization property of local field
theories in finite geometries, which justifies the finite scaling laws for
critical systems in finite geometries, and it has two very important
consequences:
{\it (i)} The scaling hypothesis at the basis of the direct renormalization
\index{renormalization!direct}
approach, which relies explicitly on calculations with finite manifolds, is
shown to be valid to all orders in perturbation theory;
{\it (ii)} for ``abstract" manifolds with dimension $D<2$, with the only
exception of open polymers ($D=1$ open surface), the following hyperscaling relation
\cite{r:Dup87}\ relating the configuration and the $\nu$ exponents holds:
\begin{equation}
\gamma\ =\ 1\,-\,\nu\,d\ \ .
\label{eHypSc}
\end{equation}

\section{Self-avoiding Manifold at the tricritical $\Theta$-point:}
\begin{figure}[t]
\centerline{\epsfbox{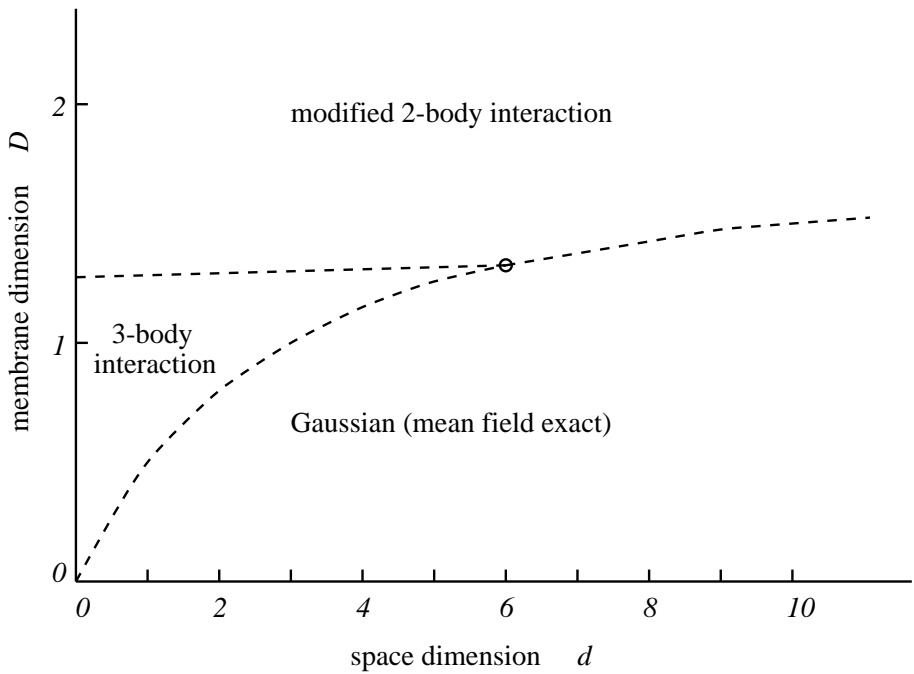}}
\label{fdDTheta}
\caption{Relevant interactions for the $\Theta$-point in the $d$-$D$ plane.}
\end{figure}
Finally, let me briefly discuss recent results obtained with K. Wiese on the
scaling behavior of polymerized membranes at the $\Theta$-point \cite{r:WD95}.
This point separates the swollen phase,
where the self-avoidance repulsive forces that I considered previously
dominate, from the dense collapsed phase, where short ranged attractive forces
dominate.
At the $\Theta$-point the effective two body repulsive coupling $b$ vanishes,
and two different interactions may become relevant.
The first one is the 3-body contact repulsion, which is usually considered for
polymers
\begin{equation}
\raisebox{-6.ex}{\epsfbox{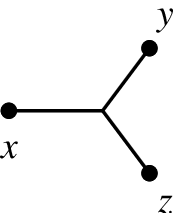}}\  
\ =\ \int\!\int\!\int\,d^D\!x\ d^D\!y\ d^D\!z\ 
\delta^d\big(\vec r(x)-\vec r(y)\big)\,\delta^d\big(\vec r(x)-\vec r(z)\big)
\ .
\label{f3Int} 
\end{equation}
The second one is a modified 2-body interaction, repulsive at short range but
attractive at larger range ($\Delta_{\vec r}$ is the $d$-dimensional Laplacian)
\begin{equation}
\raisebox{-2.ex}{\epsfbox{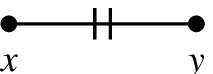}}\  
\ =\ -\ \int\!\int\,d^D\!x\,d^D\!y\ \Delta_{\vec r} \,
\delta^d\big(\vec r(x)-\vec r(y)\big)\ .
\label{f2'Int}
\end{equation}
Calculations at first order are not feasible analytically, and already require
numerical evaluations of complicated integrals.
The results of such one loop calculations are schematically depicted on
Fig.~\ref{fdDTheta}, where the domains where the 3-body and modified
2-body terms are respectively relevant are shown.
This indicates that the last modified 2-body term is the relevant one for
2-dimensional manifold in any external dimension $d$.
There is also a quite interesting and non-trivial crossover between the two
terms around $D=4/3$ $d=6$, which must be studied by a double
$\epsilon$-expansion.

\section{Conclusion:}
The theoretical study of the scaling behavior of polymerized flexible membranes
leads to the development of new multilocal continuum field theories, and to new
applications of renormalization group methods.
I hope that these methods will lead to a quantitative progress in the
understanding of the behavior of real 2-dimensional polymerized membranes.
This requires results beyond first order in the
$\epsilon$-expansion (recall that $D=2$ correspond to $\epsilon=4$), and
a better understanding of the relation between this RG approach and more
heuristic or approximate methods, such as variational methods or approximate
recursion relations.
The sophisticated renormalization theory for multilocal models presented here
should
hopefully also find  applications in other problems of statistical physics,
or in other areas of theoretical physics.

\end{document}